\documentclass[prl,twocolumn,aps,superscriptaddress,showpacs]{revtex4}
\usepackage{graphicx}
\usepackage{epsfig}
\usepackage{bm,hyperref}

\def\be{\begin{equation}}
\def\ee{\end{equation}}
\def\ba{\begin{eqnarray}}
\def\ea{\end{eqnarray}}
\def\hmf{H_{\rm s}}

\begin{document}
\title
{Anomalous Monopole In an Interacting Boson System}
\author{Biao Wu}
%\author{Biao Wu(Îâì­)}
%\email{bwu@aphy.iphy.ac.cn}
\affiliation{Institute of Physics,
Chinese Academy of Sciences, Beijing 100080, China}
%\author{Qi Zhang(ÕÅÆð)}
\author{Qi Zhang}
\affiliation{Department of Physics and Center for Computational
Science and Engineering, National University of Singapore, 117542, 
Republic of Singapore}
\affiliation{Institute of Physics, Chinese Academy of Sciences,
Beijing 100080, China}
%\author{Jie Liu(Áõ½Ü)}
\author{Jie Liu}
\affiliation{Institute of Applied Physics and Computational
Mathematics, Beijing 100088, China}
\date{Feburary 19th, 2008}
\begin{abstract}
Anomalous monopole of disk shape is found to exist in the semiclassical theory
of a two-mode interacting boson system. The quantum origin of this anomaly 
is the collapsing or bundling of field lines of Berry curvature caused by
the interaction between bosons in the semiclassical
limit. The significance of this anomalous monopole
is twofold: (1) it signals the failure of the von Neumann-Wigner theorem 
in the semiclassical limit; (2) it indicates a breakdown of  
the correspondence principle between quantum and classical dynamics.
\end{abstract}
%03.65.Sq Semiclassical theories and applications
%03.65.Vf Phases: geometric; dynamic or topological
%05.45.Mt Quantum chaos; semiclassical methods
%03.75.-b Matter waves
\maketitle

Magnetic monopole was first discussed by Dirac as a 
quantization condition for electric charge\cite{Dirac1931PRS}.  
Although it has never been observed in experiment as an 
fundamental particle, the monopole has fascinated physicists 
ever since\cite{Patrizii2003arXiv}.  
Interestingly, monopoles have attracted great attention 
in a very different context as degeneracies or diabolical 
points of energy levels in parameter 
space\cite{Wilkinson1984PRS,Berry1984PRS,Avron1989CMP}.  
The examples include the degeneracy of Bloch bands in the Brillouin 
zone\cite{Fang2003SCI} and energy levels in molecular 
magnets\cite{Garg2001PRB,Bruno2006PRL}. 
The monopoles in this context are found to be crucial to 
understanding these systems.

In this work we study a two-mode interacting boson system that depends
on three external parameters. For simplicity, we focus
on its ground state, which is doubly degenerate at one isolated point 
in the parameter space. We find that the field lines of 
Berry curvature emanating from the point are curved due to 
the interaction between bosons. Moreover, at large $N$ limit, 
that is, when the number of bosons $N$ increases to infinity,  
the field lines collapse and bundle into a two-dimensional disk 
whose radius is determined by the interaction strength. 

At large $N$ limit, this boson system can be well described
by a mean-field theory\cite{Leggett2001RMP,Wu2006PRL}. 
Since this boson system belongs to a class of quantum systems
which become classical at large $N$ limit\cite{Yaffe1982RMP}, 
this mean-field can be regarded as a semiclassical theory. 
We discover that the semiclassical (or mean-field) ground state 
of this boson system is degenerate at every point on the 
two-dimensional disk mentioned above. This means that the
whole disk is a monopole. This is in stark contrast with what 
is demanded by the von Neumann-Wigner theorem\cite{Wigner1929PHYZ}:
the monopole in a three-dimensional parameter space is always a 
point-like object. Therefore, this anomalous monopole of disk shape
indicates that the von Neumann-Wigner theorem fails in the semiclassical
limit. Our further analysis shows that the magnetic charge 
is not uniformly distributed in the disk while its total charge is
still $2\pi$, the Chern number\cite{Avron1989CMP}.
In addition, this anomalous monopole is compared to an
anomalous monopole that is formed in a trivial fashion.

The Berry curvatures are computed for this system within
the semiclassical theory and compared to the results in
the quantum description. The matching becomes better
as $N$ increases as expected from the correspondence principle
between quantum and classical dynamics\cite{Berry1985JPA}. 
However, on the monopole disk, the Berry curvature differs
significantly between its semiclassical result and quantum
result even in the large $N$ limit. This shows that 
the existence of the anomalous monopole 
indicates a breakdown of the correspondence principle. 
This breakdown is analyzed from a fresh perspective by regarding the
three external parameters as the dynamical variables
of a massive classical particle. 

The two-mode interacting boson system  is described by 
the following second quantized  Hamiltonian 
\begin{eqnarray}
\label{quantized}
\hat{H}_{N}&=&\frac{X}{2}(\hat{a}^{\dag}\hat{b}+\hat{a}\hat{b}^{\dag})+
\frac{iY}{2}(\hat{a}\hat{b}^{\dag}-\hat{a}^{\dag}\hat{b})+\nonumber\\
&&\frac{Z}{2}(\hat{a}^{\dag}\hat{a}-\hat{b}^{\dag}\hat{b})
-\frac{\lambda}{4V}(\hat{a}^{\dag}\hat{a}-\hat{b}^{\dag}\hat{b})^{2},
\label{second}
\end{eqnarray}
where $\hat{a}^{\dag}$, $\hat{a}$ and $\hat{b}^{\dag}$, $\hat{b}$
are bosonic operators for two different quantum states,
respectively, $\lambda>0$ is the interacting strength between
bosons, and $V$ is the volume of the system. 
The three parameters, $X$,$Y$, and $Z$, characterize the
influence from environment or another system. This Hamiltonian 
has its root in modeling the Bose-Einstein condensates in a 
double-well potential\cite{Wu2006PRL}. It also belongs to
a class of Hamiltonians studied in Refs.\cite{Garg2001PRB,Bruno2006PRL} 
for single molecule magnet if we introduce 
$\hat{J}_x=(\hat{a}^\dag\hat{b}+\hat{b}^\dag\hat{a})/2$,
$\hat{J}_y=i(\hat{a}^\dag\hat{b}-\hat{b}^\dag\hat{a})/2$,
and $\hat{J}_z=(\hat{a}^\dag\hat{a}-\hat{b}^\dag\hat{b})/2$. 
The focus of studies in Refs.\cite{Garg2001PRB,Bruno2006PRL} is
on the pattern and topological property of the monopoles.
In this work we examine the ``magnetic'' fields, i.e., 
Berry curvatures, generated by the monopoles and
their behavior in the semiclassical limit $N\rightarrow\infty$.
Note that large $N$ limit
is always taken by keeping $N/V$ constant.

For simplicity, we concentrate on the ground state of this system.
At point $X=Y=Z=0$, we have $\hat{H}_N=-\lambda
(\hat{a}^{\dag}\hat{a}-\hat{b}^{\dag}\hat{b})^2/4V$, whose
ground state is either $\langle\hat{a}^{\dag}\hat{a}\rangle=N$
or $\langle\hat{b}^{\dag}\hat{b}\rangle=N$. This means
that the ground state of this boson system is doubly degenerate
at point $X=Y=Z=0$. The ground state
is not degenerate elsewhere in the parameter space. 

\begin{figure}[!tb]
\begin{center}
\includegraphics[bb=15 15 285 235,width=7.5cm]{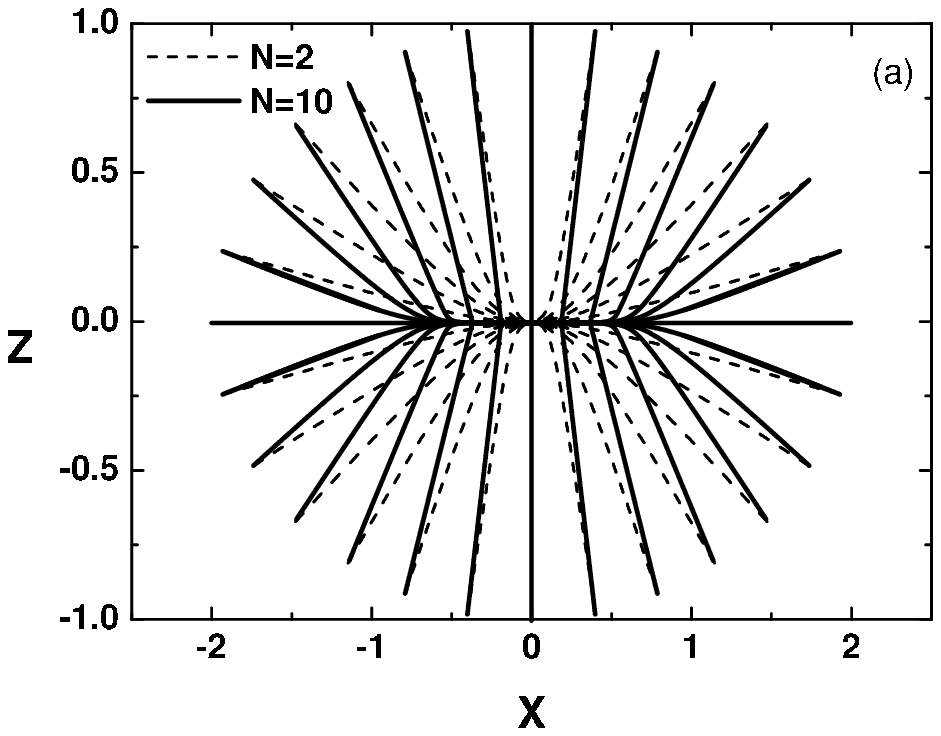}
\includegraphics[bb=15 15 285 235,width=7.5cm]{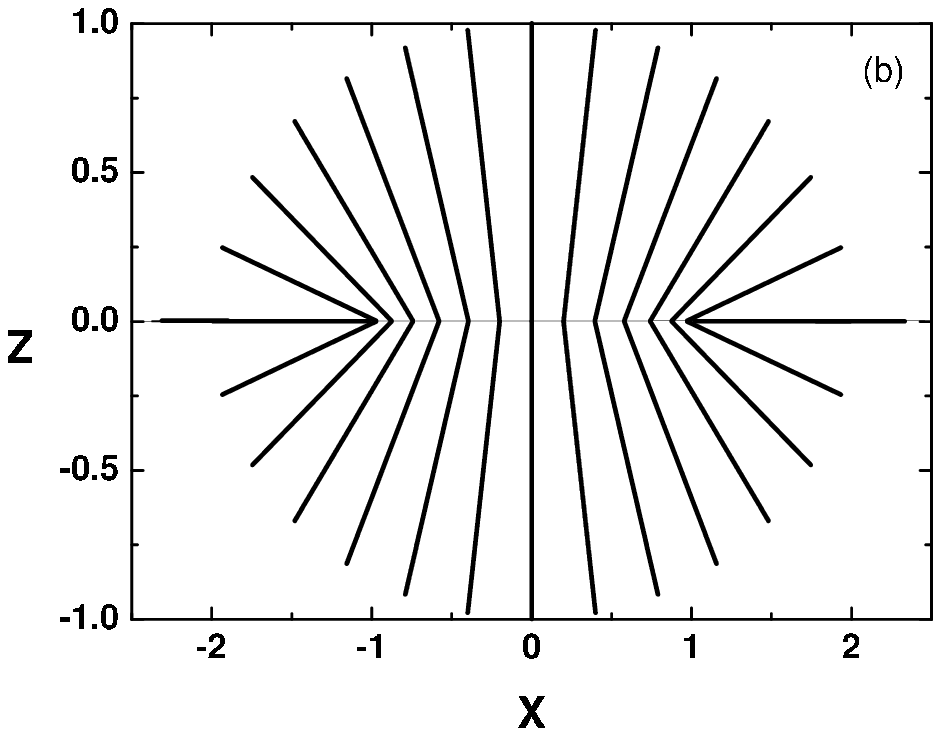}
\end{center}
\caption{Field lines of Berry curvature near the
monopole for (a) the second-quantized Hamiltonian (dashed lines are
for $N=2$ and solid lines are for $N=10$); (b) the mean-field
Hamiltonian. Due to the symmetry around $Z$-axis, the $Y$
component is omitted. $c=1$. The lines in (b) are not 
straight as they appear.} 
\label{mline}
\end{figure}
 
This degenerate point or monopole at $X=Y=Z=0$ generates ``magnetic'' 
field ${\bf B}_{N}$ (Berry curvature) in the parameter
space spanned by $X,Y,Z$. As one usually uses field lines to illustrate
a magnetic field, we have computed numerically the field lines of Berry
curvature and plotted them in Fig.\ref{mline}(a). For clarity, only
the results for $N=2$ and $N=10$ are plotted. Nevertheless an 
interesting trend is clearly demonstrated: the field lines
are curved towards a disk defined by $\sqrt{X^2+Y^2}=c=N\lambda/V$ 
and $Z=0$; the curving gets stronger as $N$ increases. In fact, 
our numerical results show that the field lines will collapse 
and bundle (or converge) into the disk when $N$ approaches infinity. 
As we know, a magnetic monopole (or an electric charge) can be 
viewed as the converging point or the emitting source of  
field lines. This collapsing (or converging) of field lines 
suggests that the whole disk become a monopole in the limit 
$N\rightarrow\infty$. This is indeed the case as we shall show.

At large $N$ limit, this boson system becomes ``classical'' and can 
be described by the following mean-field (or semiclassical) 
Hamiltonian\cite{Leggett2001RMP,Wu2006PRL},
\begin{eqnarray}
\label{mean-field}
\hmf&=&\lim_{N\rightarrow\infty}\frac{\hat{H}_{N}}{N}
=\frac{X}{2}(a^{*}b+ab^{*})+\frac{iY}{2}(ab^{*}-a^{*}b)\nonumber\\
&&+\frac{Z}{2}(|a|^{2}-|b|^{2})
-\frac{c}{4}(|a|^2-|b|^2)^{2},
\end{eqnarray}
where $a$ and $b$ are complex amplitudes for the system in
the two quantum modes. The normalization is one, i.e.,
$|a|^2+|b|^2=1$. This kind of nonlinear Hamiltonian also
appears in photoassociation systems\cite{Pu2007PRL,Itin2007PRL}.

\begin{figure}[!htb]
\begin{center}
\includegraphics[width=6.0cm]{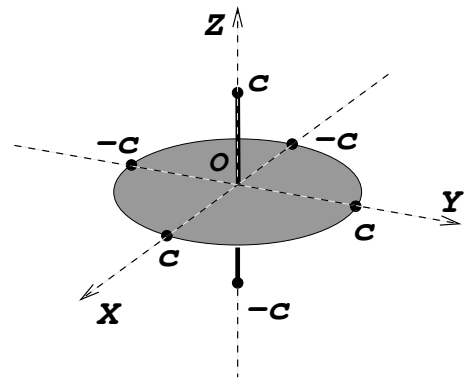}
\end{center}
\caption{Anomalous monopoles. The disk is the monopole for the
ground state and the thick vertical line is the monopole 
for the highest eigenstate.}
\label{monopole}
\end{figure}

Within this semiclassical description, the ground state of this
system is given by
\begin{equation}\label{i}
|\phi\rangle\equiv\left(\begin{array}{c}a\\b\end{array}\right)
=\left(\begin{array}{c}\sqrt{\frac{1-p}{2}}\\
-\sqrt{\frac{1+p}{2}}\frac{X+iY}{\sqrt{X^{2}+Y^{2}}}\end{array}\right),
\end{equation}
where  $p$ is the solution of the following equation,
\begin{equation}\label{h}
p\sqrt{X^{2}+Y^{2}}=(Z+cp)\sqrt{1-p^2}.
\end{equation}
This equation has one real root when $\sqrt{X^{2}+Y^{2}}\ge c$.
When $X^{2}+Y^{2}<c^2$, it can have three real roots. In
particular, when $Z=0$, two of the three real roots given by
$p=\pm\sqrt{1-(X^2+Y^2)/c^2}$ have the same energy
and are for the ground states. This means that in the semiclassical
description of the system, the ground state is degenerate on
the disk given by $X^2+Y^2<c^2$ and $Z=0$. In other 
words, the whole disk is a monopole(see Fig.\ref{monopole}). 

This anomalous disk-shaped monopole is very surprising.
According to the von Neumann-Wigner theorem, the 
accidental degeneracy of a quantum system occurs only at isolated 
points in a three dimensional parameter space\cite{Wigner1929PHYZ}.
In other words, the monopole must be a point. This is indeed 
the case when our system is described by the second-quantized 
Hamiltonian: the ground state is degenerate
only at point $X=Y=Z=0$ as we have already pointed out. 
However, in the semiclassical description,
that is, at large $N$ limit, the monopole is a two-dimensional
disk as the result of the curving and collapsing of field lines.
This shows by example that the von Neumann-Wigner theorem does not hold
in the semiclassical limit.

The significance of this anomalous monopole can be appreciated
from a different angle. We consider the highest eigenstate
of this system. For this eigenstate, its semiclassical monopole is 
a line as shown in Fig.\ref{monopole}. However, this line-shaped 
monopole which appears also anomalous is trivial, not as significant as 
the disk-shaped monopole. The reason is as follows. In the second 
quantized model, the highest eigenstate has $N$ equally spaced degenerate 
points along $Z$ axis between $-c$ and $c$. As $N$ approaches infinity, 
these degenerate points merge into a line, forming 
in a trivial fashion the one dimensional monopole.

Let us now examine this disk-shaped anomalous monopole in detail.
Although the semiclassical Hamiltonian $\hmf$ is nonlinear,  
the Berry curvature $\bm{\mathcal B}$ of this monopole can 
be computed as in a linear system\cite{Wu2005PRL}. That is
to compute the curl of the vector potential 
${\bf A}=\langle\phi|\nabla|\phi\rangle$ 
with $|\phi\rangle$ given in Eq.(\ref{i}). 
The Berry curvature $\bm{\mathcal B}$ is found to be 
\begin{equation}\label{curvature}
\bm{\mathcal B}=\frac{p^{3}}{2(cp+Z)^{2}(cp^{3}+Z)}({\bf
R}+cp\hat{z}),
\end{equation}
where ${\bf R}=\{X,Y,Z\}$ and $\hat{z}$ is the unit vector for
$Z$ direction. This result
is plotted as field lines in Fig.\ref{mline}(b). It is
apparent that these semiclassical field lines away from
the monopole disk are very similar to the field lines 
obtained with the second quantized model. Note that
$\bm{\mathcal B}$ has two different values on the monopole disk
due to the double degeneracy of the ground state.
By integrating $\bm{\mathcal B}$ over a closed surface 
around a small area in the disk, we find that the ``magnetic'' 
charge is not uniformly distributed over the disk. The charge 
distribution is 
\begin{equation}
\rho=\frac{1}{c\sqrt{c^2-(X^{2}+Y^{2})}}. \label{density}
\end{equation}
%This analytical result is consistent with our numerical computation. 
The integration of this charge density over the whole disk
gives us a Chern number of $2\pi$. So, although the monopole has 
changed from a point to a disk as the semiclassical limit is
approached, the total charge does not change. It is worthwhile to 
mention that the total charge of the line-shaped monopole is infinite
as easily inferred from its trivial origin.

Berry\cite{Berry1985JPA} once established a semiclassical relation 
between Berry phase\cite{Berry1984PRS} and Hannay's 
angle\cite{Hannay1985JPA,Liu1998PRL} 
in accordance with the correspondence principle.  
This semiclassical relation basically says that the two-forms
for Berry phase and Hannay's angle (the two-form for Berry
phase is the usual Berry curvature) are the same in the semiclassical
limit $\hbar\rightarrow 0$. This semiclassical relation should
hold in this interacting boson system. We
define two pairs of conjugate variables,
$p_{a}=\sqrt{i\hbar}a^{*}$, $q_{a}=\sqrt{i\hbar}a$ and
$p_{b}=\sqrt{i\hbar}b^{*}$, $q_{b}=\sqrt{i\hbar}b$ for the
semiclassical Hamiltonian\cite{Zhw2006PRL}. The quantization is
realized with  the following commutators,
\begin{equation}
[\hat{q}_{a},\hat{p}_{a}]=[\hat{q}_{b},\hat{p}_{b}]=i\hbar/N.
\end{equation}
One can obtain the second quantized Hamiltonian(\ref{quantized})
with the following substitution $\hat{a}=\sqrt{N/i\hbar}\hat{q}_{a}$,  
$\hat{a}^\dagger=\sqrt{N/i\hbar}\hat{p}_{a}$ and 
$\hat{b}=\sqrt{N/i\hbar}\hat{q}_{b}$, $\hat{b}^\dagger=\sqrt{N/i\hbar}\hat{p}_{b}$.
These commutators show why $N\rightarrow\infty$ is 
the semiclassical limit. As a result, the semiclassical
relation established by Berry\cite{Berry1985JPA} for this boson system is
\begin{equation} 
\label{phaserelation}
\lim_{N\rightarrow\infty}\delta{\bf
B}=\lim_{N\rightarrow\infty}(\frac{{\bf B}_{N}}{N}-\bm{\mathcal
B})=0\,,
\end{equation}
Note that the Hannay's angle in the semiclassical system $\hmf$ is
just the Berry phase generalized for nonlinear quantum system 
in Ref.\cite{Wu2005PRL} and 
the semiclassical Berry curvature $\bm{\mathcal B}$ is
the two-form for this Hannay's angle. 

\begin{figure}[!htb]
\begin{center}
\includegraphics[width=7.0cm]{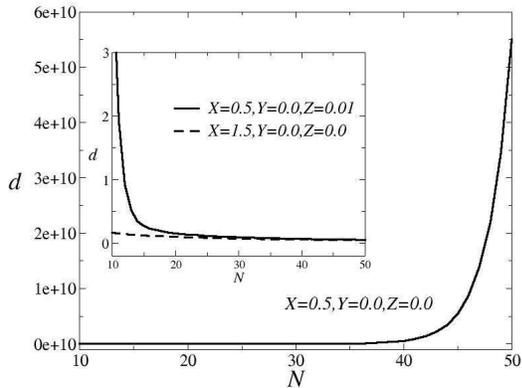}
\end{center}
\caption{The difference $d=|\delta{\bf B}^{l}|$ as
a function of the number of bosons $N$ at a point on the monopole
disk. The inset shows the results for points away from
the disk. $c=1$.}
\label{divergence}
\end{figure}

Our numerical results show that the relation (\ref{phaserelation}) indeed
holds almost everywhere in the parameter space except on
the monopole disk. On the disk, the semiclassical Berry curvature
$\bm{\mathcal B}$ has a non-zero $\hat{z}$ component while the
quantum ${\bf B}_N$ always points radially in the $Z=0$ plane. 
Furthermore, the quantum Berry curvature ${\bf B}_N$ diverges 
exponentially with $N$ on the
monopole disk while the in-plane component of the semiclassical
$\bm{\mathcal B}$ does not. We define $d=|\delta{\bf B}^{l}|$, where 
the superscript $l$ denotes the component of the vector 
parallel to the $XY$ plane. The difference $d$ is plotted 
in Fig.\ref{divergence}, where we see the
difference $d$ increases exponentially with $N$.
This diverging difference shows that the semiclassical relation
in Eq.(\ref{phaserelation}) is broken. Therefore, the disk-shaped
monopole also signifies the breakdown of the corresponding
principle between quantum mechanics and classical mechanics. 
In the following, we shall look into
this breakdown from a very different angle, showing that
some quantum effect remains in the semiclassical limit.

%%%%%%%%%%%%%%%%%%%%%%%%%%%%%%%%%%%%%%%%%%%

We treat the three parameters, ${\bf R}=\{X,Y,Z\}$, as dynamical 
variables of a massive and classical particle, whose Hamiltonian 
is $H_c={\bf P}^2/2M+U({\bf R})$. In this way we obtain a 
Born-Oppenheimer type system where a fast quantum system is coupled 
to a heavy and slow classical system\cite{Zhw2006PRL}
\begin{equation}
H=\langle\Psi|\hat{H}_{N}|\Psi\rangle+H_c\,,
\label{couple1}
\end{equation}
where $|\Psi\rangle$ is a general wavevector for the boson system.
It is reasonable to expect that at the large $N$ limit we can simplify
the above Hamiltonian by replacing its first part by the semiclassical 
Hamiltonian in Eq.(\ref{mean-field}),
\begin{equation}
H^\prime=N\hmf+H_c\,. 
\label{couple2}
\end{equation}
However, in the following, we shall show that the two Hamiltonians
$H$ and $H^\prime$ do not have the same dynamics even in the limit
$N\rightarrow\infty$. That is, there is a difference between
the two Hamiltonians which does not vanish as $N$
increases.

For a Born-Oppenheimer system, it is well known that the
dynamics of the slow system $H_c$ is controlled by two forces,
Born-Oppenheimer force and geometric
force\cite{Zhw2006PRL,Robbins1993PRS1,Robbins1993PRS2}. The 
Born-Oppenheimer force ${\bf
F}_{\rm BO}=-\nabla E_n({\bf R})$ is the gradient of an eigenenergy of
the fast system. If ${\bf B}_{g}$ is
the Berry curvature of the fast system, the dynamics of the slow
system is then given by
\begin{equation}
\label{slow} 
M\ddot{\bf R}={\bf F}_{BO}+\dot{\bf R}\times{\bf
B}_g-\nabla U({\bf R})\,.
\end{equation}
One can use  either the second quantized Hamiltonian $\hat{H}_N$
or the semiclassical Hamiltonian $\hmf$ to compute both forces.

We consider a special case, where the slow classical particle is set with 
the initial condition, $X=r<c$, $Y=0$, $Z=0$, and $\dot{X}=0$, $\dot{Y}=v$ 
and $\dot{Z}=0$ while the fast boson system is kept in its ground
state. We set up ourselves to a task to design a potential $U({\bf R})$ 
so that the slow particle stays in $Z=0$ plane and makes a circular motion.
When $N$ is large, one would feel safe to design $U({\bf R})$ by
using the semiclassical Hamiltonian $H^\prime$ to 
compute ${\bf F}_{BO}$ and ${\bf B}_{g}$.
However, due to the exponential breakdown of Eq.(\ref{phaserelation})
discussed above, such designed $U({\bf R})$ will not be able to keep 
the classical particle in the $Z=0$ plane: the strong parallel component
of ${\bf B}_N$ will kick the particle out of the plane.
This shows that there is always some physical consequence which
can not be counted in the semiclassical theory. 
It is interesting
note that Berry and Robbins once pointed out that a kind of friction
in a chaotic classical system does not exist in its corresponding quantum
system and called it discordance\cite{Robbins1993PRS2}. What
we observe here is similar to this discordance although our system
is not chaotic in the semiclassical limit.

In conclusion, we have found an anomalous monopole of disk shape in 
a two-mode interacting boson system. This kind
of anomalous monopole should exit in a general interacting
boson system. For example, if one manages to compute
the Bloch bands of an interacting boson system in a three
dimensional periodic
potential, one should expect such an anomalous monopole
in the Brillouin zone. We have further demonstrated that this 
anomalous monopole is an indication of the failure of the 
von Neumann-Wigner theorem in the semiclassical limit
and the breakdown of the correspondence principle 
between quantum and classical dynamics.

We thank the helpful discussion with Junren Shi and Qian Niu. This
work is supported by the ``BaiRen'' program of Chinese Academy of
Sciences, the NSF of China (10504040,10725521), and the 973 project of
China(2005CB724500,2006CB921400).

%\bibliography{/home/bwu/references/berry}
%\bibliographystyle{apsrev}

\end{document}